\documentstyle[emulateapj]{article}

\def\etal{{\rm et~al.\ }}

\def\kms{{\rm \;km\;s^{-1}}}

\def\kmsmpc{\kms\;{\rm Mpc}^{-1}}

\newcommand{\PSbox}[3]{\mbox{\rule{0in}{#3}\includegraphics{#1}\hspace{#2}}}

\lefthead{Croft, \etal} 
\righthead{Simulating the Effects of Intergalactic Grey Dust}

\begin{document}

\title{ Simulating the effects of intergalactic grey dust }

\author{
Rupert A.C. Croft$^{1},$  
Romeel Dav\'{e}$^{2}$,
Lars Hernquist$^{1}$, and Neal Katz$^{3}$
} 
 
\footnotetext[1]{Harvard-Smithsonian Center for Astrophysics, 
Cambridge, MA 02138; rcroft,lars@cfa.harvard.edu}
\footnotetext[2]{Princeton University Observatory, Princeton, NJ 08544;
 rad@astro.princeton.edu}
\footnotetext[3]{Department of Physics and Astronomy, 
University of Massachusetts, Amherst, MA, 01003;
nsk@kaka.phast.umass.edu}

\begin{abstract} 

Using a high-resolution cosmological hydrodynamic simulation, we
present a method to constrain extinction due to intergalactic grey dust
based on the observed magnitudes of distant Type IA supernovae. We
apply several simple prescriptions to relate the intergalactic dust
density to the gas density in the simulation, thereby obtaining dust
extinctions that may be directly compared to the observed distribution
of supernova magnitudes.  Our analysis is sensitive to the spatial
distribution of grey dust, but is not dependent on its intrinsic
properties such as its opacity or grain size.  We present an
application of our technique to the supernova data of Perlmutter et
al., who find that their high redshift sample is $\sim 0.2$ magnitudes
fainter than the expectation for a non-accelerating, low-density
 universe. We find that for grey dust to be responsible, 
it must be distributed quite
smoothly, e.g., tracing intergalactic gas. More realistic dust
distributions, such as dust tracing the metal density, are inconsistent
with observations at the $1.5-2 \sigma$ level.  Upcoming
observations and improved modelling of the dust distribution should lead
to stronger constraints on intergalactic grey dust extinction.

\end{abstract}
 
 
\section{Introduction}

Recent observations of Type IA Supernovae (SNe) at redshifts up to
$z\sim 0.8$ (\cite{ri98}; Perlmutter et al. 1999, hereafter
\cite{per99}) have made possible classical cosmological tests that
require standard candles, such as the magnitude-redshift relation.  The
most dramatic result is that these SNe appear dimmer (by $\sim 0.2$ magnitudes)
at high redshift than would be predicted in a
non-accelerating universe, suggesting at face value that we live in an
accelerating universe.  However, other explanations are possible,
including the one we consider here, namely that distant SNe are dimmer due to
extinction by intergalactic dust.

As distant standard candles, SNe are sensitive probes of extinction in
the intergalactic medium (IGM).  The distribution of matter in the IGM
may now be modelled accurately in the context of modern cosmology using
hydrodynamic simulations (see e.g., \cite{cen94}; \cite{her96};
\cite{dav99}).  The resulting IGM is not smooth, but rather traces
large-scale structure.  If such structure contains not only dark matter
and gas but also dust, this would result in significant variations in
SNe brightnesses due to intervening extinction.  Observationally, the
distribution of Type Ia SNe magnitudes has a very small dispersion
(\cite{per99}).  Thus by comparing simulations to the distribution of
observed brightnesses, we can set limits on the amount of dust
extinction and possibly constrain its spatial distribution with respect
to intergalactic gas.  In this {\it Letter} we present a technique for
doing this, and apply it to the SNe observations of \cite{per99}.

Intergalactic grey dust has been examined in a series of papers by
Aguirre (1999a,b; hereafter \cite{agu99}) and \cite{agu99c}, who develop
a scenario in which small grains are preferentially destroyed during
ejection from galaxies, polluting the IGM with large dust grains that
are effectively grey in the bandpasses of the SNe data.  This greyness is
necessary in order not to violate tight limits on reddening from SNe data,
which imply that galactic-type dust would provide negligible absorption
(\cite{per99}).  Furthermore, a significant fraction of the dust must
reside in the IGM.  If the grey dust causing extinction were
 present only
in the ISM of the supernova host galaxy, this would introduce too large
a dispersion in observed SNe magnitudes (\cite{ri98}).  In this study, we
assume that grey dust blends smoothly from galaxies into the surrounding
IGM.  Our analysis is insensitive to the intrinsic properties of the
dust, such as its opacity and grain size, since we use the simulations
to directly translate the observed SNe magnitude distribution into a dust
extinction in magnitudes.  It is, however, sensitive to the way in which
dust traces the distribution of gas in the IGM, and we will consider
several simple but plausible variations of this relation.

In \S~\ref{sec: sims} we describe our simulations of the IGM, and
of dust extinction.  In \S~\ref{sec: anal} we describe our analysis
method and results, including constraints on grey dust afforded by
current SN observations.  In \S~\ref{sec: disc} we discuss
systematic uncertainties, and the implications of our results.

\section{Simulations of Intergalactic Dust}\label{sec: sims}

We employ a hydrodynamic simulation of a $\Lambda$-dominated cold
dark matter model, with $\Omega_m=0.4$, $\Omega_\Lambda=0.6$,
$\Omega_b=0.02h^{-2}$, $H_0=65 \kmsmpc$, and $\sigma_8=0.8$.
Our simulation volume is $50 h^{-1}$~Mpc with $10 h^{-1}$~kpc spatial
resolution, having $144^3$ dark matter and $144^3$ gas particles, and was
evolved from $z=49\rightarrow 0$ using Parallel TreeSPH (\cite{ddh97}).

In order to obtain dust column densities along lines of sight, we consider
three different ways that dust may trace intergalactic hydrogen gas:
\begin{enumerate}
\item $\rho_{\rm dust}\propto \rho_{\rm gas}$: Dust traces gas linearly.
\item $\rho_{\rm dust}\propto \rho_{\rm metal}$: Dust traces metals linearly.
\item $\rho_{\rm dust}\propto \rho^2_{\rm gas}$: Dust traces gas quadratically.
\end{enumerate}
As our simulation makes no direct prediction for the metallicity of gas,
we adopt a heuristic prescription (c.f. \cite{cen99}), in the second
case above.  We assume that the metallicity is $10^{-2}$ solar if the
gas overdensity is less that 10, solar if the overdensity is greater
than 1000, and log-linear in between.

\PSbox{dustmap1.ps angle=-90 voffset=130 hoffset=-50 vscale=42 hscale=42}
{3.0in}{1.7in} 
\PSbox{dustmap2.ps angle=-90 voffset=130 hoffset=-40 vscale=42 hscale=42}
{3.0in}{1.7in} 
{\\\small Figure~1: Dust extinction maps in
$2.2^\circ\times 2.2^\circ$ patches of sky
out to $z=0.05$ (left) and $z=0.5$ (right), for
(a) $\rho_{\rm dust}\propto \rho_{\rm gas}$ and
(b) $\rho_{\rm dust}\propto \rho^2_{\rm gas}$
The median extinction magnitude
to $z=0.5$ is set equal to $0.4$ for both (a) and (b).
The maps correspond to what would be seen against a white background.
\vskip0.1in
}
\label{fig: maps}

To extract dust extinction values from the simulations, we assume that
the gas associated with each particle is spread over its SPH smoothing
volume (see e.g., \cite{her89}).  We perform a numerical integration of
gas column density along 5000 rays cast through these volumes, at the same
time applying one of the three transformations given above to relate gas
to dust densities.  To reach the required path lengths out to $z\sim 0.5$,
we follow rays through 26 simulation volumes, each ray entering through a
random point on a random face.  This yields the column density of dust to
$z=0.5$ along each line of sight.  

In Figure~1 we show extinction maps of $2.2^\circ\times 2.2^\circ$
patches of sky, for (a) $\rho_{\rm dust}\propto \rho_{\rm gas}$ and (b)
$\rho_{\rm dust}\propto \rho^2_{\rm gas}$.  The median extinction to
$z=0.5$ was set to be equal (to $0.4$ mag) for cases (a)
and (b).  In Figure~2, we show how the mean dust extinction and its
dispersion varies with redshift.  Here, the dispersion in all three
panels was set to be the same small value, equal to the difference in
quadrature of the dispersion in SNe magnitudes at high
($\sigma_{z=0.5}=0.157$) and low ($\sigma_{z=0.05}=0.154$) redshifts
observed by \cite{per99}, namely 0.03 mag. Figs.~1 and 2 show
that the mean extinction is much greater when the dust is more smoothly
distributed.  We will now quantitatively explore the constraints that
can be put on the dust extinction by using the observed distribution of
SNe magnitudes.

\PSbox{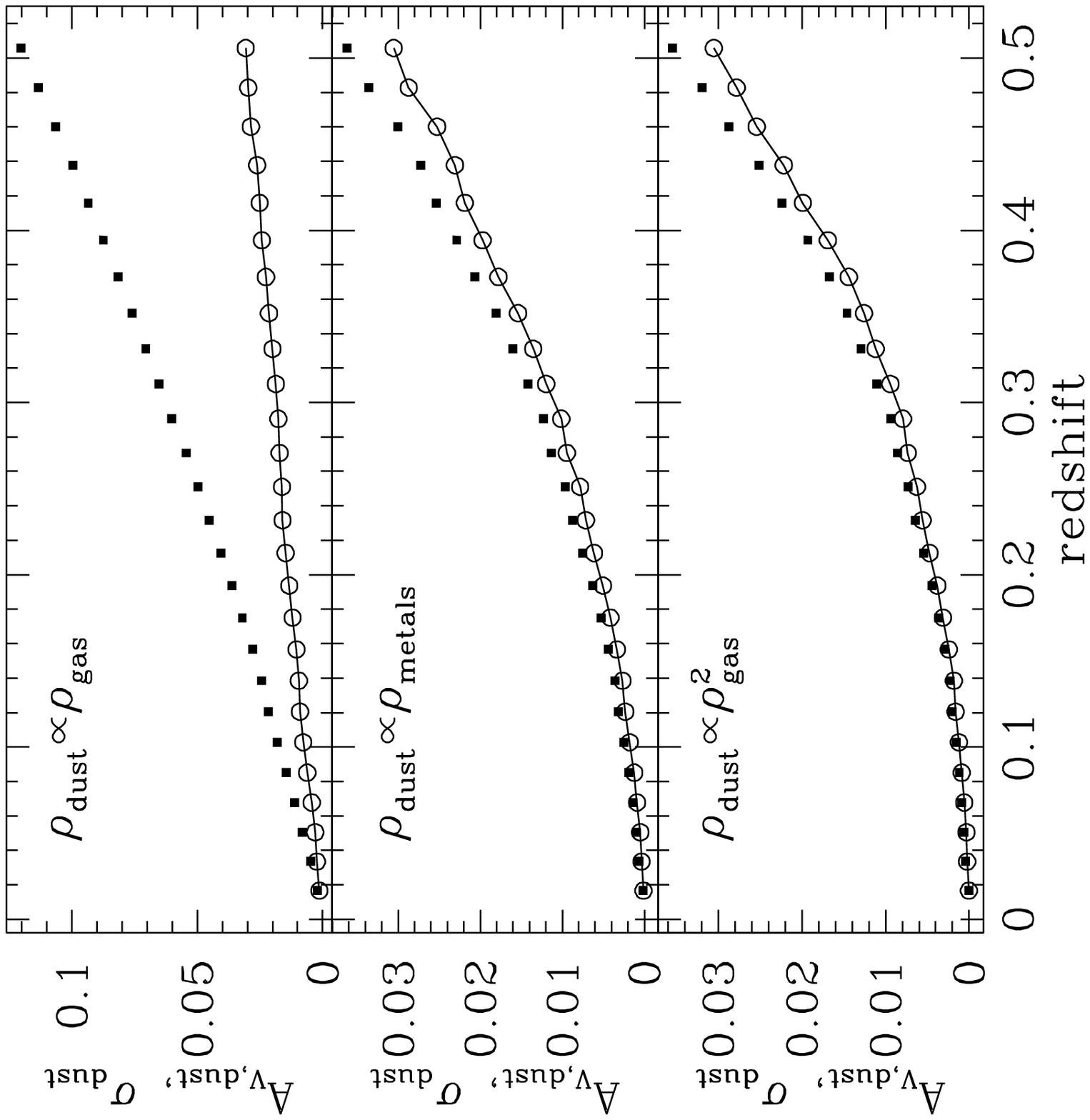 angle=-90 voffset=240 hoffset=-50 
vscale=42 hscale=42}{3.0in}{2.8in} 
{\\\small Figure~2:
Mean and dispersion of the dust extinction as a function of redshift.  
Filled squares show mean, open circles with line show the dispersion.
Values have been scaled to a dispersion of 0.03 magnitudes at $z=0.5$.
\vskip0.1in
}
\label{fig: dispz}

\section{Comparison with observations}\label{sec: anal}

We make use of two characteristics of the observed SN data
in our comparison, the change with redshift of the
dispersion in SNe magnitudes, and the shape of the histogram of SNe
magnitudes. As mentioned above, \cite{per99} found little difference in the
dispersions of two samples with $\bar{z}\sim0.05$ and
$\bar{z}\sim0.5$.  As they stated, this leaves little room for
dispersion due to dust, as this dispersion is expected to increase for
longer path lengths (see Fig.~2).  In order to quantify this, and
the effect of the distribution shape, we 
generate simulated SNe magnitudes, and compare them to the
\cite{per99} data using a maximum likelihood approach.

The observational datasets we use are both taken from \cite{per99}
(their Tables~1 and 2), being the high-$z$ SNe of the Supernova
Cosmology Project, and the low-$z$ sample of Cal\'{a}n-Tololo
SNe survey (\cite{ham96}).  We use 40 (non-reddened SNe)
 of the former SNe, 
 between $z=0.172$ and $z=0.83$ ($\bar{z}\sim 0.5$) , and 16
of the latter SNe which lie 
between $z=0.02$ and $z=0.101$ ($\bar{z}\sim 0.05$) .

We generate simulated datasets for each dust model described in \S~2.
For each dust model, we vary two parameters; first, $M_{C}$, a cosmological
magnitude shift applied to all simulated SNe at a given $z$,
normalized so that $M_{C}=0$ at $z=0.5$ corresponds to the best fitting
cosmology found by \cite{per99} (with $\Omega_{m}=0.28$ and
$\Omega_{\Lambda}=0.72$); and second, $A_{V}$, the median V-band magnitude of
dust extinction out to $z=0.5$\footnote{ We use the median in order to
be less sensitive to long tails of the distribution (c.f.
Figure~3) when renormalizing extinction values.}.
$M_{C}\approx -0.2$ then corresponds to an open model with
$\Omega_{m}\sim0.3$, and $M_{C}\approx -0.4$ to an $\Omega_{m}=1$ model.
 We generate the simulated
datasets as follows: \\
(1) We renormalize the 5000 simulated lines-of-sight
so that the median dust extinction to $z=0.5$ equals $A_{V}$.\\
(2) We add the cosmological shift, $M_{C}$.\\
(3) We broaden the magnitude distribution, which involves convolving it 
separately with a Gaussian of width given by the 
observational error of each SN, and
then adding these distributions together.  When doing this, we include an
``intrinsic'' SN dispersion of $\sigma_{\rm int}=0.17$ mag (\cite{per99}).
Varying this (by $\pm0.1$ mag) makes little difference 
to the results (see \S 4).\\
(4) We truncate the distribution at a magnitude
difference $\Delta_{M}=1$, to roughly account for the fact that
SNe along lines of sight passing through high extinction regions
would not make it into these magnitude-limited samples.

\PSbox{pdf.ps angle=-90 voffset=232 hoffset=-45 vscale=42 hscale=42}
{3.0in}{3.05in} 
{\\\small Figure~3:
The PDF of magnitude differences, $\Delta_{M}$, for 5000 simulated SNe.
The shaded region
is included in the calculation of the dispersion; non-shaded region
indicates heavily obscured SNe that are not
included in the PDF normalization.  The histogram
shows 40 of the high-$z$ SNe of \cite{per99}.
\vskip0.1in
}
\label{fig: los}

We derive the probability distribution function (PDF) of SNe magnitude
differences $\Delta_{M}$ predicted by the simulation, so that the
predicted number of SNe between  $\Delta_{M}$ and $\Delta_{M}+d\Delta_{M}$
is $N P(\Delta_{M})d\Delta_{M}$, where $N$ is the number of observed
SNe. Figure~3 shows histograms of the PDF of $\Delta_{M}$,
for $A_{V}=0.4$ mag and $M_{C}=-0.4$. We also plot the observational
data of \cite{per99}. In effect, for this plot, we have brightened
the simulated SNe to mimic an $\Omega_{m}=1$ model and then dimmed
them with dust. We can see that in panel (a), where the dust is fairly
smoothly distributed, the simulated PDF is not too different from the
observations. In the other panels, which have clumpier dust, there is
a skewness not seen in the observed data, due to a long tail of high
extinction lines-of-sight.

For each set of simulated lines of sight, we form the relative likelihood
of drawing all the observed SNe magnitudes,
${\cal{L}}=\prod_{i}^{N}P(\Delta_{M,i})$,
where $\Delta_{M,i}$ is the magnitude difference of SN $i$. We define
the quantity $S=-2\ln{\cal{L}}$ and assume that $S$ follows a $\chi^{2}$
distribution in order to derive confidence limits on the parameters
$M_{C}$ and $A_{v}$.  In the present analysis, we use results at two
different redshifts ($z=0.05$ and $z=0.5$), and combine the two by adding
the values of $S$. It is this step, combining the likelihoods
at two different redshifts, which constrains the
amount of additional dispersion (or the change
in the shape of the magnitude distribution) 
due to grey dust between low
and high redshifts.  Contours of $\Delta S=2.3, 6.2$, and $11.8$
(the difference in $S$ from its minimum value), representing $1, 2 $
and $3 \sigma$ intervals of joint confidence, are plotted in Figure~4.

\PSbox{mlikcont.ps angle=-90 voffset=280 hoffset=-80 vscale=50 hscale=50}
{3.0in}{3.20in} 
{\\\small Figure~4:
1, 2 and 3$\sigma$ contours for the cosmological magnitude shift,
$M_{C}$, with respect to a model with
$\Omega_{\Lambda}=0.72$ and $\Omega_{M}=0.28$, and the median extinction
due to grey dust ($A_{V}$).
\vskip0.1in
} 
\label{fig: mlcontour}

Figure~4 shows
that the smoothest distribution of dust easily accomodates
enough extinction to reconcile a non-accelerating ($M_C=-0.2$) or flat
($M_C=-0.4$) universe, as there is a strong degeneracy between the
cosmological magnitude shift and dust extinction.  The other cases,
where dust traces metals and $\rho^2_{\rm gas}$, are coincidentally
quite similar. Such a distribution of dust would 
 be mildly disfavored in a non-accelerating open universe 
 and ruled out at $\sim 99\%$ confidence in 
an Einstein-de-Sitter universe.
  The constraints arise mostly due to the shape of
the distribution; the fact that the observed dispersions are similar
at $z=0.05$ and $z=0.5$ is relatively unimportant, only making a
noticeable difference in panel (a) where the shape is similar to the
observed distribution.

\section{Discussion}\label{sec: disc}

We have found that current SNe datasets appear to have some power to
constrain grey dust models.  We find it somewhat unlikely (at the 
$1.5-2$ level) that sufficient grey dust could be distributed in the
relatively clumpy fashion expected for the metal distribution (c.f.,
\cite{gne97}, \cite{dav98}, \cite{cen99}) to reconcile the \cite{per99}
data with a non-accelerating universe.  This 
suggests that any substantial grey dust component must be largely
segregated from the galaxies where it was formed.  In the specific grey
dust model of \cite{agu99}, sputtering does destroy dust more
effectively in denser regions, but once in the IGM the large grains have
long lifetimes, so that dust is still likely to trace the metals.
The question of how the dust and metals are distributed can be answered
self-consistently by modelling the relevant physical processes 
(dust and metal ejection from galaxies)
directly in the simulations (e.g., Aguirre et al., in preparation).

There are many alternative explanations for SNe appearing dimmer in the
past.  For example, there may be metallicity effects in the host galaxy
(\cite{hof99}), intrinsic evolution (\cite{rie99}), observational
selection differences between CCDs used for distant SNe and
photographic plates used for nearby samples (\cite{how99}), or time
evolution of the gravitational constant (\cite{ame99}; \cite{gar99}).
Such effects would alter the interpretation of our parameter $M_{C}$.
Gravitational lensing magnification (\cite{met99}) would
also change the magnitude dispersion at higher redshifts.

The study we have presented is reasonably general, in that our analysis
is not dependent on the (unknown) microscopic properties of this
hypothetical intergalactic dust.  Still, there are some possible
systematic uncertainties, which we now consider.  We assume (as do
\cite{per99}) that the observed dispersion in SNe magnitudes in excess
of the estimated observational errors is an intrinsic property of SNe
and does not vary with redshift.  One could envision scenarios in which
the intrinsic dispersion is lower at high redshift, thus allowing more
dispersion from dust. However, as mentioned earlier, most of the statistical
power of our analysis comes from the {\it shape} of the distributions,
which is not significantly affected by a change in the intrinsic
dispersion. Also, we find that when we decrease the intrinsic
SN dispersion ($\sigma_{\rm int}$), models with more dust fit slightly
better at high-$z$, but the low-$z$ fit becomes worse, a trade-off which 
means that the overall results hardly change.
The distribution shape depends on
our assumption that the intrinsic dispersion has a Gaussian
distribution in magnitudes, whereas a distribution skewed to fainter
magnitudes could weaken our constraints.
With a larger sample of low-z SNe we might be in a position to 
test this, by using the distribution of low-z SN magnitudes to 
make a simulated high-z sample with the correct intrinsic distribution shape. 
Also, we have made a simplifying approximation in our simulated
high-z samples,
by using dust extinction magnitudes that result from integrating
the dust contribution from $z=0$ to exactly $z=0.5$. The real
SNe are at a range of redshifts; our constraints are effectively
conservative because of this.  We have also assumed that the total mass
of dust does not change with redshift from $z=0.5$ to $z=0$; dust
increasing with time would strengthen our constraints, while a decrease
seems implausible.  Finally, the truncation of our simulated magnitude
distribution at $+1$ mag is a rather approximate procedure.
We have tried changing the cutoff, and find that with none
(an unrealistic case) the constraints become stronger, as we might
expect. With a lower cutoff, the right
side of the observed distribution is not reproduced.
One additional degree of freedom would involve changing
the functional form of the cutoff.
In the future, as observations improve, we plan to simulate such observational
selection effects more carefully.

On the simulation side, our relatively low mass resolution ($m_{\rm
gas}=8.5 \times 10^8 M_{\odot}$) means that we miss fluctuations in the
extinction that occur on smaller mass scales.  If we had higher
resolution, this should have the effect of making our dust constraints
tighter. We find that the dispersion in projected magnitudes
largely depends on {\it small} scale fluctuations.  We tested this by
splitting the simulation into small sub-volumes and then shuffling them
to remove large scale correlations (see e.g., \cite{bl88}) before
projection, and we obtained similar results.  Changing the assumed
cosmological model could have some effects, although it is
difficult to conceive of models which have less power on
small scales while fitting other constraints (see e.g.,
\cite{wc2000}).

The models of \cite{agu99} have grains that do produce some reddening,
typically in the infrared.  A complementary approach to ours is
therefore to use information from different color bands. Such an
approach was used by \cite{per99}, for example, to constrain normal
galactic-type dust.  Recent near-IR observations by \cite{ri00} have
made reddening constraints tight even for non-standard large dust
grains.  Our approach does not make any use of color information, and
so constrains the most extreme scenario in which the dust is totally
``grey''.  

In summary, we have used cosmological hydrodynamic simulations to
explore how intergalactic grey dust could affect observations of high
redshift supernovae, and how supernova data can constrain grey dust
extinction.  We conclude that only a fairly smooth
distribution of dust could readily mimic the effect of an accelerating
Universe.  Such a distribution would be strange, as the dust 
would be strongly biased away from metal producing regions.  More
realistic dust distributions are mildly disfavored, but upcoming
samples of SNe (at current redshifts) should enable us to put tighter
limits as they more precisely determining the shape of the brightness
distribution. At higher redshifts ($z\ga 1$), grey dust 
predicts that SNe should show increased dimming, while
decreased dimming would be strong evidence for a
cosmological constant.

\bigskip
\acknowledgments
We thank Anthony Aguirre, Bruce Draine, and Bob Kirshner for useful 
discussions, and David Weinberg for helpful comments on the manuscript.

\end{document}